\documentclass[preprint]{sigplanconf}

\usepackage{hyperref}
\usepackage{amsmath,amsthm}
\usepackage{hypernat}
\usepackage{xcolor}
\usepackage[utf8]{inputenc}
\usepackage{listings}

\usepackage{tikz}
\usepackage[skins,listings]{tcolorbox}

\tcbset{listing engine=listings}
\tcbset{listingbox/.style={%
    enhanced,
    boxsep=-9pt,
    left=15pt,
    arc=2pt,
    boxrule=1pt,
  }
}

\definecolor{green}{RGB}{0, 180, 0}

\lstdefinestyle{custompython}{%
    belowcaptionskip=1\baselineskip,
    breaklines=true,
    frame=none,
    xleftmargin=\parindent,
    language=Python,
    showstringspaces=false,
    basicstyle=\footnotesize\ttfamily,
    keywordstyle=\bfseries\color{green!40!black},
    commentstyle=\itshape\color{purple},
    identifierstyle=\color{black},
    stringstyle=\color{green},
    keywords=[2]{as,True,False},
    keywordstyle=[2]\bfseries\color{green!40!black},
    keywords=[3]{np,sp,plt},
    keywordstyle=[3]\bfseries\color{blue},
    numbers=none,
    columns=fullflexible,
}

\newtcblisting{mylisting}{%
  listingbox,
  listing only,
  listing options={style=custompython},
}

\def\loopy{Loo.py}

\newtheorem{loopyrule}{Rule}

\begin{document}


\setlength{\pdfpageheight}{\paperheight}
\setlength{\pdfpagewidth}{\paperwidth}

\conferenceinfo{ARRAY14}{June 9--10??, 2014, Edinburgh, United Kingdom}
\copyrightyear{2014}
\copyrightdata{978-1-nnnn-nnnn-n/yy/mm}
\doi{nnnnnnn.nnnnnnn}




\titlebanner{\loopy: Transformation-based code~generation}        
\preprintfooter{Submitted to \href{http://www.sable.mcgill.ca/array/}{ARRAY14}}   

\title{\loopy: transformation-based code~generation for GPUs and CPUs}

\authorinfo{Andreas Klöckner}
           {University of Illinois at Urbana-Champaign}
           {andreask@illinois.edu}

\maketitle

\begin{abstract}
  Today's highly heterogeneous computing landscape places a burden on programmers
  wanting to achieve high performance on a reasonably broad cross-section of machines.
  To do so, computations need to be expressed in many different but
  mathematically equivalent ways, with, in the worst case, one variant per
  target machine.

  \loopy, a programming system embedded in Python, meets this
  challenge by defining a data model for array-style computations and a
  library of transformations that operate on this model. Offering transformations
  such as loop tiling, vectorization, storage management, unrolling,
  instruction-level parallelism, change of data layout, and many more,
  it provides a convenient way to capture, parametrize, and re-unify
  the growth among code variants. Optional, deep integration with \verb|numpy|
  and PyOpenCL provides a convenient computing environment where
  the transition from
  prototype to high-performance implementation can occur in a gradual,
  machine-assisted form.
\end{abstract}

\category{D}{3}{4} --- Code generators
\category{D}{1}{3} --- Concurrent programming
\category{G}{4}{} --- Mathematical software




\section{Introduction}
\label{sec:intro}
As computer architectures and execution models diversify, the number of
mathematically equivalent ways a single computation can be expressed is
growing rapidly. Unfortunately, only very few of these program variants
achieve good machine utilization, as measured in, e.g.\ percentages of peak
memory bandwidth or floating point throughput. Optimizing compilers that,
with or without the help of user annotations, equivalently rewrite user
code into a higher-performing variant have been the standard solution to
this issue, although the goal of a compiler whose built-in optimization
passes robustly make the sometimes complicated trade-offs needed to achieve
good performance has remained somewhat elusive.

\loopy\ takes a different approach.
\loopy\ code is most often embedded in an outer controlling program in the
high-level programming language Python.
The user first specifies the computation
to be carried out in a language consisting of a tree of polyhedra
describing loop bounds along with a list of instructions, each tied to a node in
the tree of polyhedra. The specification provided by the user is
deliberately only weakly ordered, providing freedom to the code generator.

Once a computation is specified as described above, its description is held
within an object which is open to inspection and manipulation from within
the host language. These manipulations occur by applying a variety
of transformations that \loopy\ makes available. Most (but
not all) of the transformations provided by \loopy\ exactly preserve the
semantics of the specified code. This is different from the conventional
compiler approach in a number of important ways:
\begin{itemize}
  \item Intermediate representations are deliberately open and intended to
    be inspected and manipulated by the user. An advanced user can easily
    implement their own transformations, extending the library already
    available.
  \item Instructions, loop bounds, and transformations together uniquely
    specify the code to be generated. \loopy\ does not attempt to be
    intelligent or make choices on behalf of the user, all while retaining an
    interface high-level enough to be usable by moderately
    technical end users.
  \item Conventional compilers carry a considerable burden in proving that
    any rewriting they apply does not change the observable behavior of the
    program. Explicitly invoked transformations allow more flexibility.  By
    invoking a transformation, the user may assume partial responsibility
    for its correctness.  This puts changes within reach that would be
    difficult or impossible to apply with conventional compiler
    architectures, such as changes to globally visible data layouts.
  \item Unlike traditional `pragma'-type compiler directives,
    transformations are applied under the control of a full-scale
    programming language. This means that code generation can react to the
    target hardware or the workload at hand.

    In addition, control from a high-level programming environment
    encourages reuse and abstraction within the space of transformations,
    which aids users in dealing with larger-scale code generation tasks, in
    which, possibly, a large number of similar computational kernels need to
    be generated.
\end{itemize}
Once a computation has been transformed into a sufficiently high-performance
variant, the last task performed by \loopy\ is the generation of OpenCL C
kernel code. If \loopy\ is
used from within Python, and specifically, with PyOpenCL
\citep{kloeckner_pycuda_2012}, some extra convenience features are available.

PyOpenCL, much like its sister project PyCUDA, provides access to a
low-level, high-performance parallel computing environment (OpenCL) from a
high-level programming language (Python), facilitating run-time code
generation (`\emph{RTCG}'). In addition to this foundational functionality
and numerous parallel programming primitives, PyOpenCL provides an array
object that behaves much like and is intended to fill a similar role as the
popular \texttt{numpy} \citep{vanderwalt_numpy_2011} array object, with
which it tightly integrates.  If used to operate on PyOpenCL or
\texttt{numpy} array objects, loopy can automatically infer types left
unspecified in user code, facilitating generic programming. Optionally, it
will also determine the values of parameters that specify array bounds,
strides, and offsets. It does so with the help of a runtime layer
that allows fast and user-friendly invocation of generated \loopy\ kernels.

Note that \loopy\ does not \emph{require} the use of Python as the host
language for generated code, or the use of PyOpenCL for that matter.  A
few extra conveniences are available with these packages and languages. But
since \loopy\ generates OpenCL kernels, one or several of these can be
generated ahead of time (say, by a script) and used from any type of host
program.

The literature on code generation and optimization for array
languages is vast, and no attempt will be made to provide a survey of the
subject in any meaningful way. Instead, we will seek to highlight a few
approaches that have significantly influenced the thinking behind \loopy,
are particularly similar, or provide ideas for further development.
\loopy\ is heavily inspired by the polyhedral model of expressing
static-control programs \citep{feautrier_automatic_1996,bastoul_code_2004}.
While it takes significant inspiration from this approach, the details of
how a program is represented, beyond the existence of a loop domain, are
quite different.  High-performance compilation for GPUs, by now, is hardly
a new topic, and many different approaches have been used, including
ones using OpenMP-style directives
\citep{lee_openmpc_2010,han_hicuda_2011},
ones that are fully automatic \citep{yang_gpgpu_2010}
ones based on functional languages \citep{svensson_obsidian_2010}, and
ones based on the polyhedral model \citep{verdoolaege_polyhedral_2013}.
Other ones
define an automatic, array computation middleware
\citep{garg_velociraptor_2012} designed as a back-end for multiple languages,
including Python. Automatic, GPU-targeted compilers for languages embedded
in Python also abound
\citep{catanzaro_copperhead_2011,rubinsteyn_parakeet_2012,continuum_numba_2014},
most of which transform a Python AST at run-time based on various
levels of annotation and operational abstraction.

Code generators just targeting one or a few specific workloads (often
matrix-matrix multiplication) using many of the same techniques available
in \loopy\ have been presented by various authors, ranging from early work
such PhiPAC
\citep{bilmes_optimizing_1997} to more recent OpenCL- and CUDA-based work
\citep{cui_automatic_2011,matsumoto_implementing_2012}.

Perhaps the conceptually closest prior work to the approach taken by
\loopy\ is CUDA-CHiLL \citep{rudy_programming_2011}, which performs
source-to-source translation based on a set of user-controlled
transformations \citep{hall_loop_2010}. The two still are not quite
alike, using dissimilar intermediate representations, dissimilar levels of
abstraction in the description of transformations, and a dissimilar (static
vs.\ program-controlled) approach to transformation.

\section{A Tour of \loopy}
\label{sec:tour}


\loopy's capabilities are most conveniently explained by example.  It will
be thus be expedient to present a tour of \loopy's interface.
\loopy\ works at the granularity of a
(short-to-medium-length) subroutine, which, in keeping with terminology
from OpenCL, is called a \emph{kernel}. Within a kernel, \loopy\ assumes
mostly static control flow, although some forms of data-dependent control
are allowed. It is intended for
convenient expression of `number-crunching'-type computations.

\subsection{\loopy's data model}
We begin with a very simple kernel that reads in one vector, doubles it,
and writes the result to another.
\begin{tcolorbox}[listingbox]
\begin{lstlisting}[style=custompython,gobble=2]
  knl = loopy.make_kernel(
      "{ [i]: 0<=i<n }",  # loop domain
      "out[i] = 2*a[i]")  # instructions
\end{lstlisting}
\end{tcolorbox}
\noindent
The above snippet of code illustrates the main components of a \loopy\
kernel:
\begin{itemize}
  \item The \emph{loop domain}: \verb|{ [i]: 0<=i<n }|. This defines
    the integer values of the loop variables for which instructions
    (see below) will be executed.
    It is written in the syntax of the \texttt{isl} library
    \citep{verdoolaege_isl_2010}.  \loopy\ calls the loop variables
    \emph{inames}. In this case, \verb|i| is the sole iname. The loop
    domain is given as a conjunction of affine equality
    and inequality constraints. Integer divisibility constraints (resulting
    in strides) are also allowed. In the absence of divisibility
    constraints, the loop domain is convex.

    Note that \verb|n| is not an iname in the example. It is a
    \emph{parameter} that is passed to the kernel by the user. \verb|n| in this
    case determines the length of the vector being operated on.

    The user may have knowledge regarding parameters that might
    allow the generation of more efficient code. \loopy\ allows
    such information to be communicated using `assumptions'. For the
    example kernel above, one might specify
    \begin{mylisting}
assumptions="n > 0 and n mod 4 = 0"
    \end{mylisting}
    \noindent as a further parameter to \verb|make_kernel| to indicate that
    \verb|n| is positive and divisible by \verb|4|.  Like the loop domain,
    the assumptions are given in \verb|isl| syntax.

    To accommodate some data-dependent control flow, there is not actually
    a single loop domain, but rather a \emph{tree of loop domains},
    allowing more deeply nested domains to depend on inames
    introduced by domains closer to the root.
    This feature will not be explored in detail in this paper.

  \item The \emph{instructions} to be executed: \verb|out[i] = 2*a[i]|. These are scalar
  assignments between array elements, consisting of a left-hand
  side assignee and a right-hand side expression.
  Right-hand side expressions are allowed to contain the usual mathematical
  operators, calls to functions defined by OpenCL, and functions defined
  by the user outside of \loopy.

  Reductions are allowed, too, and are given as, for example:
  \begin{mylisting}
sum(k, a[i,k]*b[k,j])
  \end{mylisting}
  A programming interface exists that lets the user register custom
  functions, symbols, and reduction operations.
  %

  In addition to the textual format shown above, instructions and the
  expressions defining them can be provided to loopy in the form
  of an expression tree. \loopy\ uses an external library supplying
  expression trees that provides facilities for data interchange with,
  e.g.\ the \verb|sympy| and \verb|maxima| computer algebra systems.
  This is convenient if \loopy\ is used as the code generation stage
  for a larger system.
\end{itemize}
\loopy\ allows the user to
easily inspect its internal representation of the kernel
in plain-text form:
\begin{mylisting}
>>> print knl
------------------------------------------------------
KERNEL: loopy_kernel
------------------------------------------------------
ARGUMENTS:
a: GlobalArg, type: <runtime>, shape: (n), dim_tags: (stride:1)
n: ValueArg, type: <runtime>
out: GlobalArg, type: <runtime>, shape: (n), dim_tags: (stride:1)
------------------------------------------------------
DOMAINS:
[n] -> { [i] : i >= 0 and i <= -1 + n }
------------------------------------------------------
INSTRUCTIONS:
[i] out[i] <- 2*a[i]   # insn
------------------------------------------------------
\end{mylisting}
\noindent
This facility is particularly useful for debugging and as a learning tool.
Its usefulness is most clearly visible once \loopy's library of kernel
transformations comes into play, as both input and output of a given
transformation can be readily inspected.

It is apparent that there is quite a bit more information here than was
present in the vector-doubling kernel above. The bulk of this information originates
from defaults intended to be `reasonable'. When not reasonable, all of this
information can be overridden. Specifically, the following pieces of
information were added:
\begin{itemize}
  \item
    \verb|a| and \verb|out| have been classified as array arguments in
    global device memory.
  \item
    Bounds of the arrays \verb|a| and \verb|out| have been determined,
    based on the from the indices being accessed. (For some more
    complicated cases, user input may be required.)

    Like \verb|numpy|, loopy works on multi-dimensional arrays. \loopy\
    shares \verb|numpy|'s view of arrays and interoperates
    with it.

  \item
    In addition, each array axis has been given a default memory layout,
    described by (in this case, just one) `dimension tag'.

  \item
    \loopy\ has \emph{not} determined the types of \verb|a| and \verb|out|.
    The data type is given as \verb|<runtime>|, which means that these
    types will be determined by the data passed in when the kernel is
    invoked.  \loopy\ generates a variant of the kernel for each
    combination of types passed in. Each variant is heavily cached, both
    in memory as a readily executable kernel residing in an OpenCL context,
    and as an OpenCL `binary' on disk.
\end{itemize}

\subsection{Running a kernel}
Running the kernel defined above from within the host Python program is
straightforward:

\begin{mylisting}
evt, (out,) = knl(queue, a=x_vec_dev)
assert (out.get() == (2*x_vec_dev).get()).all()
\end{mylisting}
\noindent
This run-time feature makes use of PyOpenCL, introduced above. \verb|queue|
is expected to be a PyOpenCL \verb|CommandQueue| object corresponding to an
OpenCL command queue, and \verb|a| is expected to be a PyOpenCL device
array, a work-alike of a \verb|numpy| array that resides in OpenCL global
device memory. PyOpenCL device arrays, like \verb|numpy| arrays, include
type, shape and memory layout information.  The assertion in the code above
transfers the data back to the host (into newly created \verb|numpy|
arrays) and checks that the calculation was carried out correctly.

By setting the appropriate option, \loopy\ can be instructed to print the
generated (OpenCL C) source code, increasing user insight into the code
generation process:
\begin{mylisting}
>>> knl = loopy.set_options(knl, write_cl=True)
>>> evt, (out,) = knl(queue, a=x_vec_dev)
#define lid(N) ((int) get_local_id(N))
#define gid(N) ((int) get_group_id(N))
__kernel void __attribute__ ((reqd_work_group_size(1, 1, 1)))
loopy_kernel(__global float const *restrict a, int const n, __global float *restrict out)
{
  for (int i = 0; i <= (-1 + n); ++i)
    out[i] = 2.0f * a[i];
}
\end{mylisting}
\noindent
A few things are worth noting at this point:
First, \loopy\ has used the (run-time) type of \verb|x_vec_dev| to
specialize the kernel. Absent other information, and based on the single
assignment to \emph{out}, type inference has concluded that \verb|out|,
like \verb|a| in the input, contains single-precision floating point data.
Second, the sizing parameter \verb|n|, while technically being an argument,
did not need to be passed, as \loopy\ is able to find \verb|n| from the
shape of the input argument \verb|a|.

It is also possible to obtain just the generated OpenCL C source code,
without running a kernel or making use of any run-time features.

\subsection{Ordering}
The following example highlights \loopy's ordering semantics:
\begin{mylisting}
knl = loopy.make_kernel(
  "{ [i,j,ii,jj]: 0<=i,j,ii,jj<n }",
  """
  out[j,i] = a[i,j] {id=transpose}
  out[ii,jj] = 2*out[ii,jj] {dep=transpose}
  """)
\end{mylisting}
\noindent The purpose of this code is to compute the transpose of a
two-dimensional array and, then, as a separate operation, double each entry
in place. One way to achieve correctness of this program is to require that
the transpose be complete before the multiplication is begun.

Each instruction in a \loopy\ kernel only `sees' the subset of inames with
which it is directly concerned. Speaking more precisely, given a set of
\emph{active} inames, the \emph{projection} of the loop domain onto these
inames determines the iname values for which the instruction is executed.
Usage of these projections may dictate the order in which loops are nested.

\loopy's programming model is completely \emph{unordered} by default. This
means that: \begin{itemize} \item There are no guarantees regarding the
    order (or concurrency) with which the loop domain is traversed. In the
    previous kernel, \verb|i==3| could be reached before \verb|i==0| but
    also before (or in parallel with) \verb|i==17|. A program is only
    well-formed if it produces a valid result irrespective of this
    ordering.

  \item In addition, there is (by default) no ordering between instructions
    either.

    Ordering among instructions can be introduced by explicitly notated
    dependencies, as used above and discussed below.

  \item The nesting order of the loops implied by the domain is, by
    default, undefined.

    A requested loop nesting order can be specified in the form of a list
    indicating prioritization, whose semantics are defined as:
    \begin{loopyrule}[Loop nesting and priorities] If, during determination
      of loop nesting, an ambiguity exists (i.e. more than one iname's loop
      could be opened without affecting program semantics), prefer inames
      that occur earlier in the list of iname priorities.  \end{loopyrule}
    Note that this priority information has an advisory role only. If the
    kernel logically requires a different nesting, iname priority is
    ignored.  Priority is only considered if loop nesting is ambiguous. If
    not enough prioritization information is supplied to deduce an
    unambiguous nesting, a warning is issued.  \end{itemize} To determine
inter-instruction and inter-loop ordering, \loopy\ adheres to the following
rule: \begin{loopyrule}[Dependencies] Instruction $B$ depending on
  instruction $A$ ensures that, \emph{within} the largest \emph{shared} set
  of inames between $A$ and $B$, $A$ is executed before $B$.
\end{loopyrule} In the kernel above, two coding techniques related to this
rule can be observed. First, the doubling operation declares a
\emph{dependency} on the transpose, ensuring that it completes before the
doubling is started.  This works by giving one instruction a symbolic name
(using \verb|{id=name}|) and then referring to that symbolic name from the
list of dependencies of another instruction, (using \verb|{dep=name}|).

Second, the doubling operation uses a separate set of \verb|i|/\verb|j|
loops (\verb|ii| and \verb|jj| here). If both were part of the same
\verb|i|/\verb|j| loop nest, the dependency would only apply \emph{within}
the (shared) \verb|i|/\verb|j| loops.

Since manually notating dependencies can be cumbersome, \loopy\
additionally applies the following heuristic: \begin{loopyrule}[Dependency
  heuristic] If a variable is written by exactly one instruction, then all
  instructions reading that variable will automatically depend on the
  writing instruction.  \end{loopyrule} The intent of this heuristic is to
cover the common case of a precomputed result being stored and used many
times. Generally, these dependencies are \emph{in addition} to any manual
dependencies added via \verb|{dep=...}|.  It is possible (but rare) that
the heuristic adds undesired dependencies.  In this case, \verb|{dep=*...}|
(i.e.\ a leading asterisk) can be used to notate an \emph{exhaustive} list
of dependencies.

The set of \emph{active} inames for each instruction is either given
explicitly by the user or determined by the following heuristic:
\begin{loopyrule}[Active iname heuristic] Inames referred to by the instruction
  (say, as part of an indexing expression) are always part of the set of
  active inames.

  In addition, active inames propagate along the transitive closure of the
  depends-on relation.
\end{loopyrule}
\noindent The propagation part of the
rule is illustrated by this code snippet:
\begin{mylisting}
z = expr(iname) {id=insn0}
y = expr(z)     {id=insn1}
x = expr(y)     {id=insn2}
\end{mylisting}
\noindent In this code, \verb|insn1| will depend on
\verb|insn0|, because \verb|insn1| reads \verb|z|, whose sole writer is
\verb|insn0|.  Similarly, \verb|insn2| will depend on \verb|insn1|.
Finally, \verb|insn1| and \verb|insn2| both have \verb|iname| as part of
their active iname set, because of dependency-based propagation of active
inames.

It should be readily apparent that dependencies play a major role in the
way \loopy\ programs are specified. To help users write programs in this
manner, and to ease reasoning about this nonlinear program ordering,
\loopy\ provides a facility to visualize instruction dependencies and their
interaction with loops, through the use of the open-source
GraphViz~\citep{ellson_graphviz_2002} graph drawing tools. Example output
for the above kernel is shown below, representing a graph of dependencies
along with information on loop nesting:
\begin{center}
  \includegraphics[width=3cm]{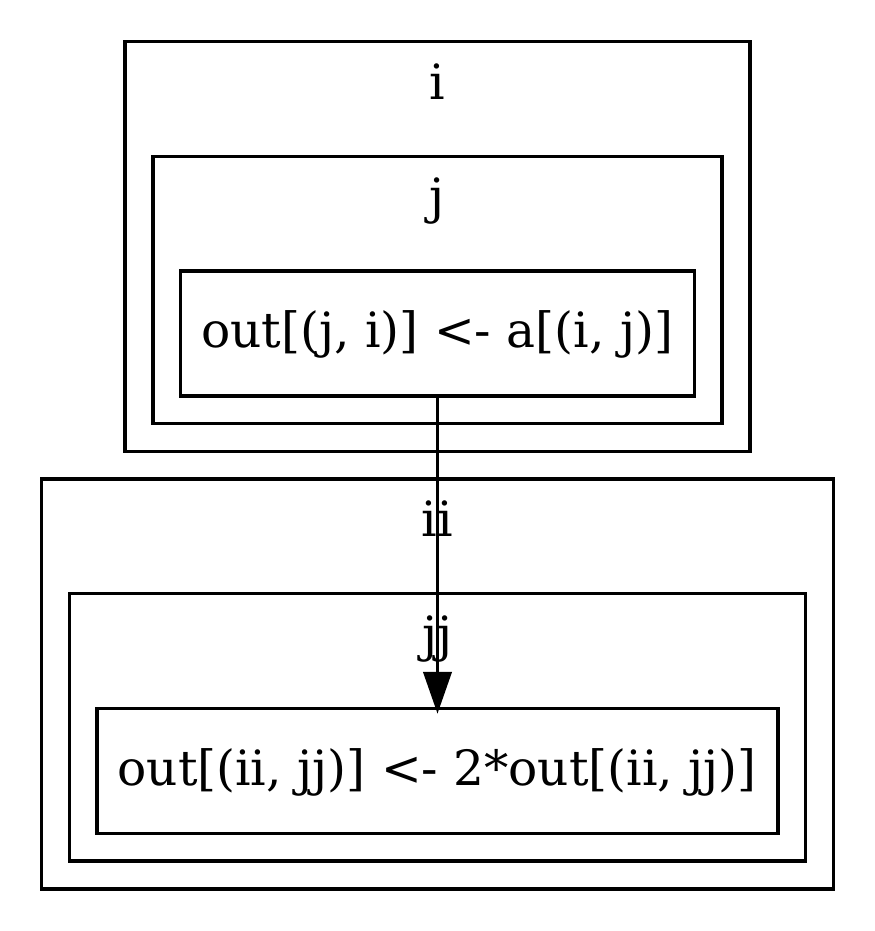}
\end{center}
\subsection{Kernel transformation}
\label{sec:transformation}
The third main pillar of \loopy\ besides the kernel data model and the
run-time integration described above is a \emph{library of transformations}
that can be applied to kernels represented in the data model. Generically,
these transformations have the following shape:
\begin{mylisting}
new_kernel = loopy.do_something(old_kernel, ...)
\end{mylisting}
\noindent
It should be noted that \loopy\ kernel instances are immutable, and thus
kernel transformations are always (at least outwardly) free from side
effects.

\subsubsection{\texttt{split\_iname} as a transformation}
As a first example, consider the \verb|split_iname| transformation, which
replaces one iname (called \texttt{OLD} below) with two new ones (called
\texttt{INNER} and \texttt{OUTER} below), one of
them of a fixed length.  The three inames are related by
\begin{equation}
  \mathtt{OLD} = \mathtt{INNER} + \mathtt{GROUP\_SIZE} * \mathtt{OUTER}.
  \label{eq:split-iname}
\end{equation}
The names \verb|INNER| and \verb|OUTER| used above have no automatic impact
on loop nesting.  The following code snippet provides an illustration:
\begin{mylisting}
knl = loopy.make_kernel(
    "{ [i]: 0<=i<n }",
    "a[i] = 0")
knl = loopy.split_iname(knl, "i", 16)
->
for (int i_outer = 0; i_outer <= (-1 + ((15 + n) / 16)); ++i_outer)
  for (int i_inner = 0; i_inner <= 15; ++i_inner)
    if ((-1 + -1 * i_inner + -16 * i_outer + n) >= 0)
      a[i_inner + i_outer * 16] = 0.0f;
\end{mylisting}
\noindent Observe that conditionals retaining the original loop bound have
been automatically introduced where necessary to maintain correctness.
\loopy\ will issue conditionals as early as possible and for groups of
instructions that are as large as feasible. Nonetheless, these conditionals
can become a serious hindrance to achieving good performance. For these
cases, an extra argument to \verb|split_iname| allows the generation of
separate code for edge and corner cases, with automatic dispatch between
edge/corner and bulk code.

To illustrate the expressiveness of the primitives supplied, note that
prioritization and the \verb|split_iname| transformation are already enough to
achieve \emph{loop tiling}.

\verb|split_iname| is implemented by first introducing the `new' inames
into the loop domain, adding the equality constraint \eqref{eq:split-iname}
and projecting out the `old' iname, and by subsequently rewriting the
instructions using \eqref{eq:split-iname} as a substitution rule. All of
these operations are available as primitives within \loopy\ or the libraries
used by it. In particular, the \verb|islpy| library \citep{kloeckner_islpy_2014},
specifically created for \loopy, provides access to all operations
implemented in \verb|isl|, providing access to advanced polyhedral
operations with relative ease.
Since \loopy's data model is documented,
it is thus feasible for users to author their own code transformations based on
it. It should be noted that while
\loopy's data model departs somewhat from the conventional polyhedral
representation of static control parts \citep{feautrier_automatic_1996},
many polyhedral ideas are nonetheless applicable and useful.

\subsubsection{Iname implementation tags}

In addition to the \verb|for| loops encountered in code examples thus far,
\loopy\ supports a number of other constructs. Each iname in loopy carries
a so-called `iname implementation tag'.

A first example of an iname implementation tag is ``\verb|unr|'', which
performs loop unrolling. The \verb|tag_inames| transformation applies
these tags.  After using \verb|split_iname| to create fixed-length sub-loop, one
might use unrolling on the fixed-length loop:
\begin{mylisting}
knl = loopy.make_kernel(
    "{ [i]: 0<=i<n }",
    "a[i] = 0",
    assumptions="n>=0 and n mod 4 = 0")
knl = loopy.split_iname(knl, "i", 4)
knl = loopy.tag_inames(knl, {"i_inner": "unr"})
->
for (int i_outer = 0; i_outer <= (-1 + ((3 + n) / 4)); ++i_outer)
{
  a[0 + i_outer * 4] = 0.0f;
  a[1 + i_outer * 4] = 0.0f;
  a[2 + i_outer * 4] = 0.0f;
  a[3 + i_outer * 4] = 0.0f;
}
\end{mylisting}
\noindent
Note that without the divisibility assumption on the vector length, more
general (but less efficient and more verbose) code would have been
generated.

Recall that \loopy\ targets the OpenCL/CUDA model of computation. Within
that model, parallelism is expressed as a two-level hierarchy of groups
of thread-like `work items', with the higher level abstracting the index
of a `processor core' and the lower level abstracting the index of a
`vector lane'. These two levels of indices are accessible by iname
implementation tags, corresponding to grid axes presented by the compute
abstraction.

Further iname implementation tags invoke instruction-level parallelism
(``\verb|ilp|''), whose main feature is duplication of work-item-level
storage to emulate multiple work items in-line in a single one, or
vectorization (``\verb|vec|'').
\subsubsection{Transforming data layout}
Mirroring the functionality of iname implementation tags, each array axis
is also associated with an implementation tag. By default an array axis
will be cast into code using a simple stride-based offset. Other options
include dispatching differing array indices to entries of explicitly
represented vectors (such as OpenCL's \verb|float4|), or to separate array
arguments altogether. This is particularly powerful because it allows
the user to switch between a structure-of-arrays and an array-of-structures
memory layout using a simple change of an array axis implementation tag.
Another set of data layout transformations available in \loopy\ introduces
padding and block granularities.
\subsubsection{Prefetching and precomputation}
An important consideration in many kinds of computational software is the
trade-off between available storage space and computational power needed to
recompute an intermediate result. \loopy\ gives the user explicit
control over this using the high-level \verb|add_prefetch| and lower-level
\verb|precompute| transformations.

\verb|precompute| is based on \emph{substitution rules} that are somewhat
like C macros. They are notated alongside instructions, but set
apart from them through the use of a \verb|:=| definition marker.
\begin{mylisting}
f(x) := x*a[x]
g(x) := 12 + f(x)
h(x) := 1 + g(x) + 20*g(x)

a[i] = h(i) * h(i)
\end{mylisting}
Before code generation, all substitution rules are expanded, leading to the
right-hand-side code in the rule being, effectively, inlined. Beyond
redundancy reduction, substitution rules serve another important purpose:
They can act as identifiers for precomputed quantities. Given an index
range covering arguments of a substitution rule, \loopy\ can allocate and
manage a temporary (OpenCL-\verb|private| or \verb|local|) array for this purpose,
and it can orchestrate any necessary synchronization necessary to ensure
precomputed data is ready when needed and not prematurely overwritten.

\verb|add_prefetch| makes use of this facility by first extracting all or
specified accesses to a given array into a substitution rule and then
relying on \verb|precompute| to generate the code loading the data.



This concludes a brief tour of some of the main features in the \loopy\
code generator and its input language. Much of \loopy's design was guided
by computational software that was built on top of it, with an eye towards
broader impact and usefulness.


\section{Experimental results}
\label{sec:experiments}
Presenting performance results for a code generator like \loopy\ is not
very meaningful in general, as the obtained performance hinges on the sequence
of transformations specified by the user to a far larger degree than
it might for an optimizing compiler. After all, \loopy\ is not a compiler,
but a code generator. Nonetheless, the argument that good performance is
achievable using \loopy's transformations merits being supported.
Table~\ref{tab:performance} summarizes performance results for a variety
of workloads across CPUs and GPUs. These performance numbers were obtained
by running \loopy's test cases against the list of devices specified
in the caption of Table~\ref{tab:performance}. Since \loopy's tests are,
for now, more focused on correctness than performance, these results
should be viewed as a lower bound, in the sense that better
performance should be available with rather limited tuning effort.

A careful exploration of how \loopy's transformation language enables
access to performance across a  variety of common numerical operations
is the subject of a forthcoming article
\citep{kloeckner_applications_2014}.

\begin{table}
  \begin{center}
  \begin{tabular}{|p{2.5cm} l|c|c|c|}
    \hline
    &  & Intel & AMD & Nvidia \\
    \hline
    \texttt{saxpy} & [GBytes/s] & 18.6 & 231.0 & 232.1 \\
    \texttt{sgemm} & [GFlops/s] & 12.3 & 492.3 & 369.4 \\
    3D Coulomb pot. & [M Pairs/s] & 231 & 10949 & 9985 \\
    dG FEM volume & [GFlops/s] & 77.4 & 1251 & 351 \\
    dG FEM surface & [GFlops/s] & 25.9 & 527 & 214 \\
    \hline
  \end{tabular}
  \end{center}
  \caption{%
    Performance results for a number of simple performance tests.
    `Intel' tests were run on an Intel(R) Core(TM) i7-3930K CPU @ 3.20GHz
    using the 64-bit Linux Intel OpenCL SDK, build 76921.
    `AMD' tests were run on an AMD Radeon HD 7990 using AMD's
    \texttt{fglrx} driver version 14.1beta1 and ICD version 14.3beta1.
    `Nvidia' tests were run on an Nvidia GeForce Titan using Nvidia's
    64-bit Linux driver and ICD version 331.49.
  }
  \label{tab:performance}
\end{table}

\section{Conclusions}
\label{sec:conclusions}

\loopy\ provides a small, modular code generation capability for
high-performance array code on CPU- and GPU-type shared memory
parallel computers. It is available under the MIT open-source
license from \url{http://mathema.tician.de/software/loopy}.

The core contributions in the approach behind \loopy\ are the following:
\textbf{(1)} A novel, partially ordered programming language and
corresponding internal representation of array-based programs based loosely
on the polyhedral model was described.
\textbf{(2)} An extensive library of transformations was presented to
act upon the internal representation that is able to capture
many commonly used tuning strategies.
\textbf{(3)} A novel way of assembling heterogeneous computational software
is presented. The approach uses a dynamic language for high-level control while
interfacing with a run-time code generator for high-performance execution.
It builds and improves upon the model of run-time code generation from a
scripting language proposed in \citep{kloeckner_pycuda_2012}.
\textbf{(4)} The data model exposes enough information for a strong
run-time interface that provides safe, efficient transitions between
host and embedded language, optionally enabling type-generic programming.
\textbf{(5)} The ideas above combine to yield good user program
maintainability by enforcing strong separation of concerns between
computation semantics and performance optimization, easily capturing
program variants and allowing optimization reuse.

While \loopy\ is a useful system \emph{today},
a number of extensions are likely to
broaden its appeal and increase its usefulness.  Improving \loopy's code
generation for reduction and adding a capability for parallel scans
\citep{blelloch_scans_1989} would unlock applications thus far out of
reach. A key feature enabling both of these as well as many more
applications is the emulation of global synchronization by mapping one
\loopy\ kernel to multiple OpenCL-style kernels. Next, while \loopy\
currently targets only OpenCL C, expanding code generation support to
further backends (such as OpenMP or CUDA C) would not only let more users
enjoy its benefits, it would also allow transform-based programming to
deliver code that is flexible with respect to device vendors and the underlying
compute abstraction.  

\loopy's kernel representation, its library of transformations, and its
runtime features combine to provide a compelling environment within which
array-shaped computations can be conveniently expressed and optimized. It
covers a number of important applications of parallel computation,
including a few of the well-known `seven dwarfs' and their extensions
\citep{asanovic_landscape_2006}. It makes great performance accessible and
the code that achieves it maintainable.


%

\acks

I would like to acknowledge tremendously influential discussions with Tim
Warburton that led to the genesis of \loopy\ and guided its design. I would
also like to acknowledge feedback from early adopters of \loopy, including
Rob Kirby, Maxim Kuznetsov, and Ivan Oseledets.

My work on \loopy\ was supported in part by
US Navy ONR grant number N00014-14-1-0117 and via a postdoctoral appointment
at NYU supported in part by the Applied Mathematical Sciences
Program of the U.S. Department of Energy under Contract DEFGO288ER25053 and
by the Office of the Assistant Secretary of Defense for Research and
Engineering and AFOSR under NSSEFF Program Award FA9550-10-1-0180.

\bibliographystyle{abbrvnat}


\bibliography{loopy}

\newpage

\end{document}